\documentclass[prc]{revtex4}

\usepackage{graphicx}
\usepackage{amsmath}
\begin{document}
\title{Universal Flow in the First Stage of Relativistic Heavy Ion Collisions}
\author{Joshua Vredevoogd and Scott Pratt}
\affiliation{Department of Physics and Astronomy,
Michigan State University\\
East Lansing, Michigan 48824}
\date{\today}

\begin{abstract}
In the first moments of a relativistic heavy ion collision explosive collective flow begins to grow before the matter has yet equilibrated. 
Here it is found that as long as the stress-energy tensor is traceless, early flow is independent of whether the matter is composed of fields or particles, equilibrated or not, or whether the stress-energy tensor is isotropic. 
This eliminates much of the uncertainty in modeling early stages of a collision.
\end{abstract}


\maketitle

\section{Introduction and basic theory}
At ultra-relativistic energies, heavy ions collide and produce enormous energy densities, exceeding several GeV/fm$^3$, well above the threshold for dissolving hadrons into partons and melting the QCD vacuum condensates \cite{Iancu:2003xm}.
Due to partial transparency the deposited energy is borne with a large collective velocity gradient along the beam axis (which will be labeled the $z$ axis). 
In the limit of infinite beam energy, the collective expansion along the $z$ axis is boost invariant \cite{Bjorken:1982qr}, with $v_z=z/t$, which corresponds to zero acceleration in the $z$ direction. 
Since the matter has no initial collective velocity in the transverse direction, transverse expansion is pressure driven with the transverse collective velocities growing until break-up at which time particles free-stream toward the detector. 
Several clear signals of the collective nature of the transverse expansion have been observed.
First, the average transverse energy of protons, kaons and pions are ordered by mass \cite{Adams:2003xp},
as expected since the collective velocity adds more energy to a more massive particle. 
Secondly, for non-central collisions the initial transverse energy density profile is anisotropic, which leads to anisotropic transverse pressure gradients and thus anisotropic flow 
The observable,
\begin{equation}
v_2\equiv\frac{1}{2}\left\langle\cos 2(\phi-\phi_{\rm r.p.})\right\rangle,
\end{equation}
quantifies the anisotropy, where $\phi_{\rm r.p.}$ is the angle of the reaction plane. 
Observed values of $v_2$ reach into the tens of percent \cite{Abelev:2008ed} and are consistent with expectations of hydrodynamic flow \cite{Kolb:2003dz}.
Finally, two particle correlations, which are measured as functions of the two momenta $p_1$ and $p_2$,  provide six-dimensional femtoscopic pictures of outgoing phase space clouds. 
The six-dimensional structure has numerous features expected from boost-invariant flow along the $z$ axis and from strong transverse flow \cite{Lisa:2005dd}.

Comparing data to simple models reveals solid evidence for the existence of strong collective flow.
However, detailed comparison with full dynamic models are required to infer quantitative information about the equation of state, viscosity, or other properties of QCD matter. 
Modeling relativistic heavy ion collisions is complicated since the matter traverses three distinct stages during the $\sim 15$ fm/$c$ expansion, each of which involves different degrees of freedom. 
In the final 5-10 fm/$c$ of the collision, thermal equilibrium is lost and hydrodynamics, even viscous hydrodynamics, is inapplicable. 
Collisions at this point are binary and microscopic simulations or Boltzmann codes can be applied with some confidence \cite{Soff:2000eh,Pratt:2008sz}. In the intermediate stage, local equilibrium is sufficiently maintained to justify viscous hydrodynamics \cite{Heinz:2005bw,Teaney:2001av}.
This is fortunate, since the quantities of greatest interest - temperature, pressure, and viscosity - form the basis of the description. 
It is during this stage that the majority of collective flow develops. 
The first stage, times less than $\lesssim 1$ fm/$c$, is the most uncertain and theoretically contentious.   
Descriptions might be based on classical QCD fields \cite{McLerran:1993ni, Teaney:2002kn}, 
or on partons \cite{Bass:2004vh}. The partons might be highly non-equilibrated with a significant fraction of the energy tied up in high-energy jets \cite{Wang:1991hta}.  
The initial transverse acceleration is driven by the conservation laws of the stress-energy tensor, $\partial_\alpha T^{\alpha\beta}=0$,   
and given the very different stress-energy tensors implied by the proposed models, one might expect very different flows to be generated during the early stage. 
The importance of understanding the early phase is underscored by noting that although the pre-hydrodynamic stage lasts only $\sim 1$ fm/$c$, 
the acceleration during that stage has a relatively higher impact on the evolution of the collision.
This is for the same reason that the start plays a relatively large role in a 100-m sprint.

The purpose of this article is to explain a non-intuitive result -- that the impulse to the collective flow provided by the initial stage is identical for a large class of simple models, 
even though the pictures yield very different evolutions of the stress energy tensor. 
The requirements for the universal behavior are threefold:
\begin{enumerate}\parskip 0pt
\item Longitudinal flow has a boost-invariant profile, $v_z=z/t$.
\item The stress-energy tensor is traceless.
\item  The anisotropy of the spatial components of the stress energy tensor is independent of the transverse coordinate and depends only on the Bjorken time $\tau\equiv\sqrt{t^2-z^2}$.
\end{enumerate}
The first assumption is reasonable for 100$A$ GeV collisions at the Relativistic Heavy Ion Collider (RHIC) and very well justified for the very high energy heavy ion collisions to be performed at the LHC.
The second assumption is well satisfied by any description based on massless particles or weakly interacting gauge fields.  
The final assumption is more subtle:  for instance, if a system suddenly changes from longitudinal fields to thermalized particles, then the anisotropy of the stress-energy tensor will also change suddenly. 
As a measure of the anistropy, one can define the quantity,
\begin{equation}
\label{eq:kappadef}
\kappa\equiv (T_{xx}+T_{yy})/2T_{00},
\end{equation}
and if $\kappa$ varies mainly with $\tau$, rather than with the transverse coordinates, the final criterion for universality is satisfied. Once the criteria for universal behavior are satisfied, the hydrodynamic stage can be initiated with known flow fields, which should depend only on the shape of the initial energy density profile. 
The simple expression for the flow fields derived here reduces, if not eliminates, the need for detailed modeling of the initial stage, and isolates the final evolution from contentious issues concerning the pre-equilibrium stage.

Collective flow is driven by the stress energy tensor, 
\begin{equation}
T^{\alpha\beta}=\left(
\begin{array}{cccc}
T_{00} & T_{x0} & 0 &0\\
T_{x0} & T_{xx} &0 &0\\
0 & 0 & T_{yy} & 0\\
0 & 0 & 0 & T_{zz}
\end{array}
\right)~,
\end{equation}
where it has been assumed that one is modeling only along the  $y=z=0$ axis, so that $v_z=0$ (although $\partial_z v_z = 1/t$), 
and $x$ refers to the radial direction with $v_y=0$ by reflection symmetry about the $yz$ plane. 
For central collisions, the initial stress-energy tensor has $ T_{xx}=T_{yy}$ and is diagonal, with $T_{xx}$, $T_{yy}$ and $T_{zz}$ representing the force per unit area pushing in the three directions.

\begin{table}
\begin{tabular}{|c|c|c|c|c|}\hline
{\bf Model} & $T_{xx}=T_{yy}$ & $T_{zz}$ & $T_{00}$ \\ \hline
Longitudinal electric field & $T_{00}$ & $-T_{00}$ & $\sim$constant\\ \hline
\parbox[c]{3.0in}{Free streaming massless particles,\newline
two-dimensional relativistic gas\newline
or fields from incoherent longitudinal currents}
 & $T_{00}/2$ & $0$ & $\sim 1/t$\\ \hline
Ideal hydrodynamics of massless gas & $T_{00}/3$ & $T_{00}/3$ & $\sim 1/t^{4/3}$\\ \hline
\end{tabular}
\caption{\label{table:Tij}
Elements of the stress energy tensor for simple models at early times.  
A wide variety of reasonable values for the stiffness of the transverse equation of state are represented by the models.
}
\end{table}
If one assumes a small velocity $v_x$ and neglects the longitudinal expansion, one finds a simple expression for the radial acceleration,
\begin{equation}
\label{eq:zerolongexp}
\frac{\partial v_x}{\partial t}=-\frac{\partial T_{xx}/\partial x}{T_{00}+T_{xx}}~.
\end{equation}
This immediately gives the impression that the transverse acceleration is driven by $T_{xx}$, which differs substantially for different pictures of the initial stage listed in Table \ref{table:Tij}. 
However, the inclusion of longitudinal expansion significantly alters the expression for the radial acceleration. 
To see this we consider the full expression for the conservation of momentum current,
\begin{equation}
\label{eq:consmom0}
\partial_t T_{0x}=-\partial_x T_{xx}-\partial_zT_{xz}-\partial_yT_{xy},
\end{equation}
Expressions for $T_{xz}$ and $T_{xy}$ for small $z$ and $y$ can be expressed using boost and rotational invariance,
\begin{equation}
T_{xz}(x,y=0,z)\approx \frac{z}{t}T_{0x},~~~T_{xy}(x,y,z=0)\approx \frac{y}{x}(T_{xx}-T_{yy}).
\end{equation}
Inserting these into the last two terms in Eq. (\ref{eq:consmom0}),
\begin{equation}
\label{eq:consmom}
\partial_t T_{0x}=-\partial_x T_{xx}-\frac{T_{0x}}{t}-\frac{(T_{xx}-T_{yy})}{x}.
\end{equation}
Similarly, conservation of the energy current,
\begin{equation}
\label{eq:conse0}
\partial_t T_{00} = -\partial_x T_{0x}-\partial_y T_{0y}-\partial_z T_{0z},
\end{equation}
can be rewritten using boost and rotational invariance to re-express $T_{0y}$ and $T_{0z}$,
\begin{equation}
T_{0z}(x,y=0,z)\approx (T_{00}+T_{zz})\frac{z}{t},~~~T_{0y}(x,y,0) \approx \frac{y}{x}T_{0x},
\end{equation}
which transforms Eq. (\ref{eq:conse0}) along the $y=z=0$ axis to
\begin{equation}
\label{eq:conse}
\partial_t T_{00} = -\frac{1}{t}(T_{00}+T_{zz})-\left(\partial_x+\frac{1}{x}\right)T_{0x}.
\end{equation}
Combining Eq.s (\ref{eq:consmom}) and (\ref{eq:conse}),
\begin{equation}
\partial_t\left(\frac{T_{0x}}{T_{00}}\right)\approx
\frac{-\partial_x T_{xx}}{T_{00}}+\frac{T_{zz}T_{0x}}{T_{00}^2t},
\end{equation}
Only early times are considered so terms of higher order in $T_{0x}/T_{00}$ and the term proportional to $(T_{xx}-T_{yy})$ are discarded. 
Since $T_{0x}/T_{00}$ rises linearly in time, one can solve for the effective acceleration $\alpha(x)$ defined by,
\begin{eqnarray}
\frac{T_{0x}}{T_{00}}&=&\alpha(x)t,~~{\rm for~small~}t,\\
\nonumber
\alpha(x)&=&-\frac{\partial_xT_{xx}}{T_{00}-T_{zz}}.
\end{eqnarray}
If the stress energy tensor is traceless and if $T_{xx}=T_{yy}=\kappa(t)T_{00}$, that is the anisotropy of $T_{ij}$ does not depend on $x$, one obtains,
\begin{eqnarray}
\alpha(x)&=&-\frac{\kappa\partial_xT_{00}}{T_{xx}+T_{yy}}\\
\nonumber
&=&-\frac{\partial_xT_{00}}{2T_{00}},
\end{eqnarray}
which is independent of $\kappa$. 
The principal result of this study,
\begin{equation}
\label{eq:universal}
\frac{T_{0x}}{T_{00}}\approx -\frac{\partial_xT_{00}}{2T_{00}}t,
\end{equation}
is surprising in that the acceleration is independent of the anisotropy of the equation of state. 
The extra strength of the transverse acceleration due to longitudinal flow is also remarkable. For the case of ideal hydrodynamics, $T_{0x}/T_{00}$ increases 50\% faster than what one would expect from  Eq. (\ref{eq:zerolongexp}), which would have been applied if there were no longitudinal expansion.

The ratio $T_{0x}/T_{00}$ can determine velocity once one has made an assumption about the equation of state and viscosity to be applied in the hydrodynamic stage. 
Combining the constraint of Eq. (\ref{eq:universal}) with the Navier-Stokes equation and knowing the initial profile for $T_{00}$ is sufficient to determine all of $T^{\alpha\beta}$. 
However, the success of this procedure depends on whether the universal flow persists to sufficiently late times so that the velocity gradient, $dv_z/dz\sim 1/t$, has subsided to the point that the Navier-Stokes conditions for the stress-energy tensor can be applied. 
Estimates for such a condition are close to $\lesssim 1$ fm/$c$. 
The goal of the next section is to exactly calculate $T_{0x}/T_{00}$ for several simple models, with widely different anisotropies of $T_{ij}$, to see whether the flow remains universal out to times near 1 fm/$c$.

Although $T_{0x}/T_{00}$ has a universal behavior for the models being considered here, other measures of flow will vary between models. 
For example, one might use the collective velocity, $u$, of the matter, i.e., the velocity for which an observer would measure $T_{0i}=0$. 
A common measure of elliptic flow \cite{Kolb:2000sd,Teaney:2001av} uses the spatial components of the stress-energy tensor,
\begin{equation}
\label{eq:epsilonpdef}
\epsilon_p\equiv \frac{\int dxdy~(T_{xx}-T_{yy})}{\int dxdy~(T_{xx}+T_{yy})},
\end{equation}
which for free particles would be proportional to $v_2$. 
All these measures are defined through the stress-energy tensor. 
The quantity representing the development of collective flow should preferable be chosen such that it is preserved during a rapid change of the microscopic degrees of freedom. Examples of such a change might be the sudden decay of longitudinal fields or the rapid isotropization of matter.
During such a transition, the basic conservation of the stress-energy tensor should remain valid, even as any of its ten independent components are allowed to change suddenly. One can consider a hyper-surface for which the microscopic form of matter is discontinuous on one side to the other. For example, for one side of the hyper-surface one might assume longitudinal fields, while applying ideal hydrodynamics on the opposite side. 
At any point on the hypersurface, one can find the four-vector $n^\alpha$ which is orthogonal to the surface. 
If $n^2=+1$ it is possible to boost to the frame where $n=(1,0,0,0)$. 
In this frame, neighboring points all undergo the transition simultaneously and the only discontinuity is in the time direction.
If one integrates the conservation equations across an infinitesimal time element, one finds the four constraints
\begin{eqnarray}
0&=&\int_{t-\delta t}^{t+\delta t} \left(\partial_t T_{0 \alpha}+\partial_i T_{i \alpha}\right)\\
\nonumber
&=&T^{\alpha 0}(x,y,z,t+\delta t)-T^{\alpha 0}(x,y,z,t-\delta t).
\end{eqnarray}
If the hyper-surface were defined by a unit four-vector where $n^2=-1$, one could boost to a frame where the discontinuity was locally static. 
In that case one would integrate the conservation equations across the surface, and if the discontinuity were in the $z$ direction, one would find:
\begin{equation}
0=T^{\alpha z}(x,y,z+\delta z,t)-T^{\alpha z}(x,y,z-\delta z,t).
\end{equation}
These are the Rankine-Hugoniot conditions for a shock wave. 
For either case, energy-momentum conservation across a discontinuity is stated as
\begin{equation}
n_\alpha T^{\alpha\beta}(x+)=n_\alpha T^{\alpha\beta}(x-).
\end{equation}

The sudden changes of state most often imposed in relativistic heavy-ion collisions, e.g., changing from fields to particles, tend to be invoked at a constant Bjorken time, $\tau=\sqrt{t^2-z^2}$, which corresponds to $n=(1,0,0,0)$ for $z=0$. 
This would force $T_{00}$ and $T_{0i}$ to remain constant across the discontinuity, and the flow as defined by $T_{0i}/T_{00}$ would be unchanged. 
In contrast, since both the flow velocity $u$ and the elliptic anisotropy $\epsilon_p$ require using the spatial components of the stress-energy tensor for their defnition, they could change instantaneously without violating energy-momentum conservation. 
Thus, these measures can provide misleading insight into whether flow has developed. 
For instance, if one considers free streaming particles, $\epsilon_p$ remains zero.
But if those particles instantaneously thermalize at a constant Bjorken $\tau$, the stress-energy tensor will instantaneously adjust itself while preserving the four spatial components $T_{0\alpha}$. 
At this point, $\epsilon_p$ suddenly becomes finite. 
Furthermore, to the degree that $T_{0i}/T_{00}$ is universal, the collective flow will be determined by the new state of matter rather than the previous state. 
For example, if one believes the matter becomes suddenly hydrodynamic at $\tau=1$ fm/$c$, all the models listed in Table \ref{table:Tij} would lead to the same flow for $\tau\ge 1$ fm/$c$, with the same collective velocities $u$ and the same $\epsilon_p$.

In the next section, simple examples are presented where the universality of flow is tested for sudden transitions to hydrodynamics. 
Similar universal behavior would apply if the transition were to viscous hydrodynamics, though the collective velocities $u$ and the elliptic anisotropy $\epsilon_p$ would be different than for transitions to ideal hydrodynamics.
Thus, once one has chosen a time $\tau_0$ at which to begin the hydrodynamic prescription, along with the viscosity coefficient, the collective velocity at $\tau_0$ is then independent of the properties of the matter assumed for $\tau<\tau_0$, as long as the three criteria listed earlier are satisfied. 

\section{Three Models}

In the previous section it was shown that for small times one finds universal behavior of the flow as defined by $T_{0x}/T_{00}$. 
In order to assess whether universal behavior applies for times of the order $\tau\lesssim 1.0$ fm/$c$, exact solutions are found for three different models:
\begin{itemize}\parindent 0pt
\item Ideal hydrodynamics of an ultra-relativistic gas with $P=\epsilon/3$.  
\item Coherent non-interacting electro-magnetic fields (field points in the same direction initially) that follow Maxwell's equations. 
This can be thought of as the field generated by two oppositely charged capacitor plates receding at $\pm c$.
\item Incoherent electromagnetic fields generated by a random ensemble of $\pm$ charges receding from one another at $\pm c$.
\end{itemize}
Each of these cases involves a traceless stress-energy tensor, 
\begin{equation}
T_{xx}+T_{yy}+T_{zz}=\epsilon,
\end{equation}
but with very different asymmetries between the initial transverse pressure $T_{xx}=T_{yy}$ and the longitudinal pressure $T_{zz}$. 
The measure of this initial anisotropy, $\kappa$, which is the ratio of the transverse pressure $T_{xx}$ to the energy density $\epsilon$ as defined in Eq. (\ref{eq:kappadef}), is 1/3 for the hydrodynamic model, 1/2 for the incoherent fields and unity for the coherent fields. 
Non-interacting massless particles would also result in $\kappa=1/2$. 
Colliding particles with insufficient cross section to thermalize would have $1/3<\kappa<1/2$. 
Thus, these three models seem to well span the range of possibilities. 
For the solutions presented in the next section, each model is initialized with the same transverse energy density profile and with zero collective velocity. 

The first model is ideal hydrodynamics with a simple equation of state $P=\epsilon/3$. 
For ideal hydrodynamics one assumes that the stress-energy tensor has the form,
\begin{equation}
\label{eq:Talphabetahydro}
T^{\alpha\beta}=(P+\epsilon)u^\alpha u^\beta-Pg^{\alpha\beta}.
\end{equation}
For this model $T_{xx}=T_{yy}=T_{00}/3$ at early times, i.e., $\kappa=1/3$. 
The model is solved numerically, beginning at $\tau=0$. 
Solving the equation of motion, $\partial_\alpha T^{\alpha\beta}=0$, is complicated by the fact that the energy density is singular at $\tau=0$. 
For that reason, the equations of motion are manipulated so that one solves for the quantities
\begin{eqnarray}
w&\equiv&\frac{T_{0x}}{T_{00}},\\
\nonumber
U&\equiv&T_{00}\tau^{4/3}.
\end{eqnarray}
The equations of motion for $w$ and $U$ are straight-forward to generate from Eq.s (\ref{eq:consmom}) and (\ref{eq:conse}), with the advantage being that the quantities have no singular behavior. 

For the second model we consider a coherent electromagnetic field generated from two oppositely charged capacitor plates receding from one another at $\pm c$. 
This is not particularly physical, as the fields are coherently pointing along the same direction, whereas in a heavy-ion collision the chromo-electric fields would be random with a coherence length set by the saturation scale 
\cite{Lappi:2006xc, Krasnitz:2002mn}. 
This model is chosen because it has the extreme $T_{xx}\approx T_{00}$ as the initial equation of state. 
For a single pair of opposite charges originating from $x=y=0$, Lienart Wiechart potentials can be used to generate the electric and magnetic fields,
\begin{eqnarray}
\label{eq:maxwellpoints}
A_z(r,t)&=&2q\int_0^\infty dt' \delta(x^2+y^2-(t-t')^2),\\
\nonumber
E_z(r,t)&=&4q\delta(r^2-t^2),\\
\nonumber
B_\phi(r,t)&=&-E_z(r,t).
\end{eqnarray}
For receeding plates with charge densities $\rho(x,y)$, one can integrate over the charges and find,
\begin{eqnarray}
E_z(x,y,t)&=&2\int d\phi~\rho(x-t\cos\phi,y-t\sin\phi),\\
\nonumber
B_x(x,y,t)&=&2\int d\phi~\rho(x-t\cos\phi,y-t\sin\phi)\sin\phi,\\
\nonumber
B_y(x,y,t)&=&-2\int d\phi~\rho(x-t\cos\phi,y-t\sin\phi)\cos\phi.
\end{eqnarray}
The fields can then be used to generate the stress-energy tensor,
\begin{eqnarray}
T_{00}&=&\frac{1}{2}(E_z^2+B_x^2+B_y^2),\\
\nonumber
T_{xx}&=&\frac{1}{2}(E_z^2-B_x^2+B_y^2),\\
\nonumber
T_{yy}&=&\frac{1}{2}(E_z^2-B_y^2+B_x^2),\\
\nonumber
T_{zz}&=&\frac{1}{2}(-E_z^2+B_x^2+B_y^2),\\
\nonumber
T_{0x}&=&-E_zB_y,\\
\nonumber
T_{0y}&=&E_zB_x.
\end{eqnarray}
The initial transverse distribution of the electric field matches the profile of the charge density, and since $T_{00}\sim E_z^2$ initially, the gaussian radius characterizing the electric field is larger than that of the resulting energy density profile by a factor of $\sqrt{2}$. 
The charge density was then chosen with the larger radius so that the resulting energy density profile would be the same as the other two models.

The third model is also based on the evolution of classical fields, but assumes that the fields resulted from a distribution of point charges $\pm q$, receeding at $\pm c$. 
The resulting fields from a single point particle are easily calculated from Eq. (\ref{eq:maxwellpoints}). 
Although the fields have a random sign, the resulting stress-energy tensor is always positive and the energy density always moves outward. 
Thus, one calculates the stress-energy tensors for a single point-charge pair, then integrates over the density of such charges to find the stress-energy tensor. 
For such a case the transverse magnetic field and the longitudinal electric fields have the same strength, which makes $T_{zz}=0$ and $\kappa=1/2$. 
Effectively, each point sends out an electromagnetic pulse which behaves exactly the same as massless partons being emitted from a point source at $z=t=0$. 
It is not surprising that the value of $\kappa=1/2$ is identical to what one would obtain from a non-interacting parton picture, where all the partons were emitted at $z=t=0$.

Complete incoherence, as assumed in the third model, is unphysical as it assumes perfectly point-like charges.
For a beam energy $E$, the  uncertainty principle precludes assigning the starting points for $z$ and $t$ to a region less than $\sim 1/E$. 
The transverse sizes of the individual charges is also limited by whatever radiative dynamics are responsible for the creation of the charge exchanges. 
In QCD this coherence length is referred to as the saturation scale \cite{Romatschke:2006nk}.
For finite-sized charges of correlation length $\lambda$, the stress-energy tensor behaves like the coherent limit for $\tau<<\lambda$ and like the incoherent limit for $\tau>>\lambda$. 
From the delta function forms for the electric and magnetic field in Eq. (\ref{eq:maxwellpoints}) one can see that the energy density, which involves squaring the fields which are described by delta functions, is infinite in the limit of $\lambda\rightarrow 0$. 
In fact the radiated density per unit rapidity behaves $\sim \alpha/\lambda$. 
While it is clear that a realistic evolution of the fields is more complicated than either the coherent or the incoherent limit even in the limit of little interaction,
if one considers an ensemble of randomly charged small but finite-sized Gaussian packets, $\kappa$ would still depend mainly on $\tau$, and not $x$ or $y$. 
The three criteria for universal early flow would still be satisfied.

\section{Results}
Here, we present the evolution of $T_{0x}/T_{00}$ for the three simple models described in Table \ref{table:Tij}, each with widely different initial anisotropies for $T_{ij}$. 
For each model, we assume a Gaussian energy density profile,
\begin{equation}
T_{00}(x,y)\propto \exp\left\{-\frac{x^2}{2R_x^2}-\frac{y^2}{2R_y^2}\right\},
\end{equation}
with radii $R_x=R_y=3$ fm for the calculations used to investigate radial flow. 
The transverse velocity and flow, $u_x$ and $T_{0x}/T_{00}$ respectively, are calculated as a function of the transverse coordinate $x$ along the $y=z=0$ axis for three early times: 0.3, 0.6 and 1.0 fm/$c$. 
Additionally, one can calculate the collective velocity that would ensue if the system were to suddenly change at those three times into pure hydrodynamic flow with a simple equation of state $P=\epsilon/3$. 
These are found by comparing to the form for the stress energy tensor for ideal hydrodynamics as defined by Eq. (\ref{eq:Talphabetahydro}). 
One solves for $u'_x$ and $\epsilon$ by matching to $T_{00}$ and $T_{0x}$,
\begin{equation}
\label{eq:T0alphamatch}
T_{00}=\frac{4}{3}\epsilon u_0^{\prime 2}-\frac{1}{3}\epsilon,~~
T_{0x}=\frac{4}{3}\epsilon u'_0u'_x.
\end{equation}
The velocity $u'_x$ will depend only on the ratio $T_{0x}/T_{00}$. Thus, if $T_{0x}/T_{00}$ exhibits universal behavior, so will $u'_x$. 
For the hydrodynamic model $u'=u$, whereas they differ for the other two models. 
From Eq. (\ref{eq:universal}), one can exactly calculate $T_{0x}/T_{00}$ to lowest order in the proper time $\tau$,
\begin{equation}
\label{eq:lineartime}
\frac{T_{0x}}{T_{00}}\approx \frac{x\tau}{2R_x^2}.
\end{equation}
Using Eq. (\ref{eq:T0alphamatch}) one can also find $u_x$ and $u'_x$ to lowest order in $\tau$,
\begin{eqnarray}
\label{eq:analyticflow}
u_x&\approx& \frac{x\tau}{(1+\kappa)R_x^2},\\
\nonumber
u'_x&\approx& \frac{x\tau}{(4/3)R_x^2}.
\end{eqnarray}

\begin{figure}
\centerline{\includegraphics[width=0.5\textwidth]{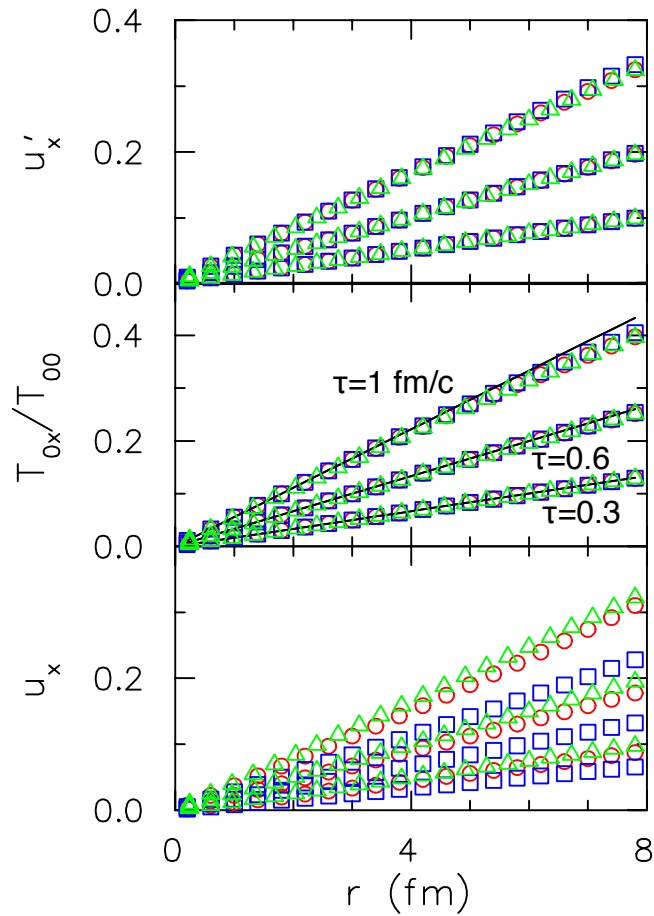}}
\caption{\label{fig:radialflow}(color online)
Lower Panel: The collective velocity profile is displayed for three models at three different times, 0.3, 0.6 and 1.0 fm/$c$. 
Ideal hydrodynamics (green triangles) has the greatest transverse radial collective flow even though it had the smallest transverse pressure, $T_{xx}$, of all three models. 
The evolution of a non-interacting coherent electric field (blue squares) had the highest pressure, but the smallest flow. 
Calculations based on electromagnetic fields arising from incoherent currents (red circles) would be the same as for non-interacting partons.\newline
Middle Panel: The flow ratio $T_{0x}/T_{00}$ is nearly universal for all three models. 
The symbols are the same as in the lower panel. 
The solid lines represent the linear approximation, $\approx\tau$, given in Eq. (\ref{eq:analyticflow}).\newline
Upper Panel: The collective velocity assuming that the matter suddenly behaves as if it were ideal hydrodynamics at the prescribed time. 
Since this ratio depends on $T_{0x}/T_{00}$, it is also nearly universal.
}\end{figure}
Figure \ref{fig:radialflow} displays $u_x$, $T_{0x}/T_{00}$ and $u'_x$ for all three models as a function of $x$ for three times, 0.3, 0.6 and 1.0 fm/$c$. 
As expected, $T_{0x}/T_{00}$ and $u'_x$ exhibit nearly universal behavior, with differences of a few percent by 1 fm/$c$. 
The linear-in-time approximation, given in Eq. (\ref{eq:lineartime}), is also displayed for $T_{0x}/T_{00}$. 
The linear approximation is remarkably effective for the first 1.0 fm/$c$.   
As expected from Eq. (\ref{eq:analyticflow}), the flow velocities, $u_x$, differ. 
What is surprising is that the models with higher values of $\kappa$ lead to lower velocities. 
This is opposite to the trend one would obtain if there were no longitudinal flow and Eq. (\ref{eq:zerolongexp}) would be have been applicable.

\begin{figure}
\centerline{\includegraphics[width=0.5\textwidth]{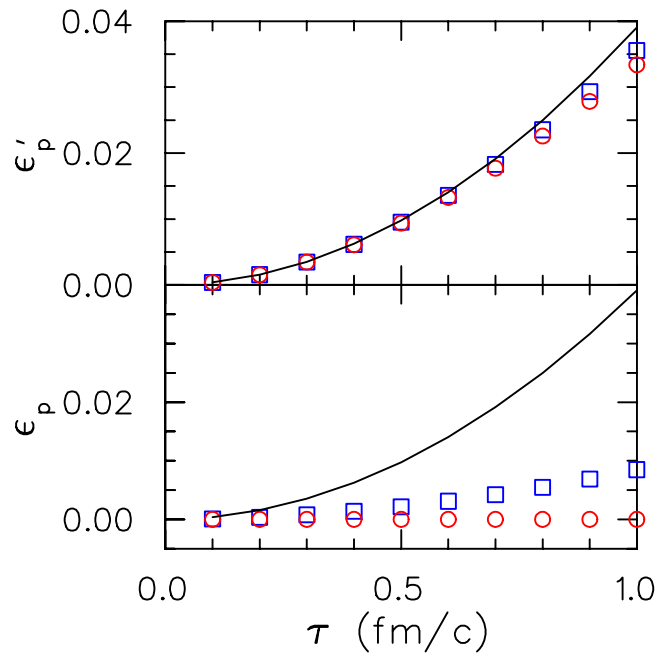}}
\caption{\label{fig:epsilonp}(color online)
Lower Panel: The elliptic anisotropy $\epsilon_p$ as defined by Eq. (\ref{eq:epsilonpdef}) for the case of coherent non-interacting initially-longitudinal electric fields (blue squares) and for fields driven by incoherent currents (red circles). 
The incoherent case yields $\epsilon_p$ is zero, exactly as one would obtain with non-interacting particles. 
The solid line shows the lowest-order (in $\tau$) quadratic contribution for ideal hydrodynamics.\newline
Upper Panel: Assuming that the matter suddenly behaves as if it were ideal hydrodynamics at time $\tau$, $\epsilon'_p$ represents the anisotropy of the altered stress-energy tensor. 
The result is close to the quadratic form approximating the behavior of ideal hydrodynamics.
}\end{figure}

Elliptic flow was evaluated by considering emission from an initial energy profile characterized by $R_x=2$, and $R_y=3$. 
As a measure of elliptic flow, $\epsilon_p$, defined in Eq. (\ref{eq:epsilonpdef}), is calculated for the two models based on coherent and incoherent fields. 
Results for the hydrodynamic model are skipped because that model was predicated on radial symmetry, although calculations have been done previously for ideal hydrodynamics with and without transverse thermalization \cite{Heinz:2002rs}.
Assuming a sudden transformation to ideal hydrodynamics at $\tau$, $\epsilon'_p$ was also calculated using the same method to calculate $u'_x$ used for the radial case above. 
Figure \ref{fig:epsilonp} shows both $\epsilon_p$ and $\epsilon'_p$ as a function of $\tau$ for both models and compares them to the small-$\tau$ expansion, $\epsilon_p\sim \tau^2$. 
The small-$\tau$ limit is found by calculating $u_x$ and $u_y$ for small times from Eq. (\ref{eq:analyticflow}) for the hydrodynamic model, $\kappa=1/3$. The collective velocities are then
\begin{equation}
u_x^{({\rm hydro})}\approx \frac{3}{4}\frac{x}{\tau}{R_x^2},~~~u_y^{({\rm hydro})}\approx \frac{3}{4}\frac{y}{\tau}{R_y^2}.
\end{equation}
Using Eq. (\ref{eq:Talphabetahydro}) for the stress-energy tensor, one can then calculate the elliptic anisotropy with some straight-forward integrals of Gaussians,
\begin{equation}
\epsilon_p^{({\rm hydro})}\approx\frac{9\tau^2}{32}\left(\frac{1}{R_x^2}-\frac{1}{R_y^2}\right).
\end{equation}
As expected, the two models agree with this simple quadratic form for $\epsilon_p'$, but differ very substantially for $\epsilon_p$. 
In fact, for the model with incoherent fields, $\epsilon_p$ remains zero for all times.
This follows from the fact that each point source contributes incoherently to the stress-energy tensor, and each point source has zero elliptic anisotropy.

Once a system has decoupled, the anisotropy $\epsilon_p$ can be equated with the angular anisotropy $v_2$,
\begin{equation}
\langle v_2\rangle\sim\frac{1}{2}\epsilon_p,
\end{equation}
where the average $\langle\cdots\rangle$ refer to an average over particles in a central rapidity bin weighted by $p_t^2/m_t$. Even though the values of $\epsilon_p$ are $\lesssim 10$\% of the $v_2$ observed experimentally, the contribution from the first fm/$c$ is substantial. 
Since $\epsilon_p$ grows quadratically in time, it is important to generate a rate of change, $d\epsilon_p/d\tau$, as quickly as possible. 
The first fm/$c$ is especially important in elliptic flow analyses for two reasons. First, one is considering non-central collisions which are smaller in overall volume and thus of shorter duration, and second, elliptic flow saturates earlier than radial flow \cite{Heinz:2002rs}.

\section{Summary}

The existence of universal flow patterns for the first $\lesssim 1.0$ fm/$c$ of a relativistic heavy ion collision has a profound impact on the modeling and interpretation of heavy ion collisions.
It eliminates many of the uncertainties plaguing the pre-thermalized stage. 
For example, if one were to use viscous hydrodynamics beginning at $\tau=1$ fm/$c$, the initial profile for $T_{0i}/T_{00}$ would be determined by the universal conditions shown here.
Given that the contribution to the final-state flow from the first fm/$c$ could be of the order of 10-20\%, it makes detailed modeling of the pre-thermalized stage unnecessary if one is only interested in the development of the evolution of the stress-energy tensor at later times.

This does not, by any means, make theoretical investigations of the pre-thermalized stage irrelevant.
Uncertainties in the shape of the initial profile would remain, including questions about the magnitude of the initial energy density and the microscopic structure.
Even though two pictures result in the same flow fields, they might have very different microscopic structure. Differing densities of quarks, gluons and kinetic temperatures should affect a variety of other observables such as electromagnetic probes \cite{vanHees:2006ng,Dusling:2006yv,Dusling:2007gi}, jet quenching \cite{Gyulassy:2003mc}, or charge balance functions \cite{Bass:2000az,Cheng:2004zy}.

\end{document}